\documentclass[aps,pre,superscriptaddress]{revtex4}
\usepackage{latexsym} 
\usepackage{times} 
\usepackage{graphicx}

\usepackage[usenames,dvipsnames]{xcolor}

\usepackage{afterpage}

\usepackage{graphicx}
\usepackage{enumerate}
\usepackage{amsmath,amsthm,amscd,amssymb}
\usepackage{mathrsfs}
\usepackage[labelfont={bf}]{caption}

\begin{document}
\title{Absence of Detailed Balance in Ecology}

\author{Jacopo~Grilli}
\affiliation{Department of Physics and Astronomy G. Galilei, Universit\`{a} di Padova, CNISM and INFN, via Marzolo 8, 35131 Padova, Italy}

\author{Sandro~Azaele}
\address{Institute of Integrative and Comparative Biology, University of Leeds, Miall Building, Leeds LS2 9JT, United Kingdom}

\author{Jayanth~R~Banavar}
\address{Department of Physics, University of Maryland, College Park, MD 20742, USA}

\author{Amos~Maritan}
\affiliation{Department of Physics and Astronomy G. Galilei, Universit\`{a} di Padova, CNISM and INFN, via Marzolo 8, 35131 Padova, Italy}

\begin{abstract}
Living systems are typically characterized by irreversible processes. A condition equivalent to the reversibility is the detailed balance, whose absence is an obstacle for analytically solving ecological models. We revisit a promising model with an elegant field-theoretic analytic solution and show that the theoretical analysis is invalid because of an implicit assumption of detailed balance. A signature of the difficulties is evident in the inconsistencies appearing in the many-point correlation functions and in the analytical formula for the species area relationship.
\end{abstract}

\maketitle

Spatially explicit models of ecology amenable to analytic solution are rare~\cite{Durrett1996,Zillio2005,Rosindell2007}. This is because ecological measures such as the species area relationship (SAR), the relation between the average number of observed species and the sampled area, crucially depend on the behavior of the correlation functions at all orders, and any truncation inevitably impairs the results. As a consequence, one needs to solve the model in full generality, a task that is highly non-trivial because stochastic theories defined on space often have stationary states for which detailed balance (DB) does not hold.

The main difference between equilibrium stationary states (those for which the DB condition holds) and non-equilibrium stationary states is reversibility. The pathways among configurations in the stationary state of an equilibrium system are symmetric under time reversal symmetry. In the stationary state of a non-equilibrium system, instead, it is possible to identify a direction of time: the 'macroscopic' history of the system with time running backwards or forwards is completely different. For instance, if DB is satisfied, the spatio-temporal patterns of declining as well as increasing species' populations should be (on average) the same, because under such a condition the decline of a species' population with time running forward is equivalent to an increase of the same population, but with the time running backward. However, there are observational findings~\cite{Bill} which show that declining species typically have sparse distributions, whereas growing ones are more aggregated. Such patterns emerge because the underlying communities are not in equilibrium and call for non-equilibrium models. Such frameworks are usually more realistic but much harder to study. Recently, O'Dwyer and Green~\cite{O'Dwyer2010} proposed a spatially explicit model (the OG model) for which the DB condition is not satisfied. Thus this model is able to show interesting irreversible phenomena, potentially in agreement with observed data. For such a model, a very interesting pattern to study is the SAR, for which there is no analytic solution based on a microscopic model incorporating fundamental processes. Remarkably, OG were able to derive an explicit formula for the SAR by using field theoretical techniques. However, their calculations were based on the incorrect assumption that DB was satisfied. The assumption is not merely a useful (and otherwise harmless) approximation, because such a simplification can result in meaningless quantities such as negative probabilities or asymmetric correlation functions.

A generic Markov process (such as the OG model) is defined via its master equation which specifies the time evolution of the probability of any configuration of the system in terms of the transition rates between any two pairs of its microscopic configurations:
\begin{equation}
\displaystyle
\frac{\partial P(C;t)}{\partial t} = \sum_{C'} \underbrace{\Bigl(W[C' \to C] P(C';t) - W[C \to C'] P(C;t) \Bigr)}_{R(C,C')} \ ,
\label{eq:mastereq}
\end{equation}
where $C$ represents a configuration of the system (e.g., in the OG model this is the set of the numbers of individuals that are present in every lattice site) and $P(C;t)$ the probability to find the system in the configuration $C$ at time $t$. The functions $W[C \to C']$ are the transition rates from the configuration $C$ to the configuration $C'$ (e.g., the birth, death and dispersal rate in the OG model).

An important property of a system is the stationary distribution of its configurations: this is given by the probability, $P_s(C)$, which makes the right hand side of eq.~\ref{eq:mastereq} equal to zero. Thus, finding $P_s(C)$ is equivalent to calculating the solution of the following equation $\sum_{C'} R(C,C') = 0$. For some simple systems, the solution can be obtained by assuming that each term in the sum is zero, i.e. $R(C,C')=0$. This equation defines the well-known DB condition. However, in general, the stationary distribution of a system is described by a probability function that does not satisfy the equation $ W[C' \to C] P_s(C') = W[C \to C'] P_s(C)$ for every $C$ and $C'$. A process whose stationary state obeys DB is said to be an equilibrium stationary state. In this case, the stationary state does not exibit any stationary current or flow (e.g. of energy or of particles) and it is possible to prove that the stationary solution obey thermodynamic principles and can be described by the powerful tools of equilibrium statistical mechanics~\cite{Zia2007}.

Unfortunately, we are not usually given the stationary distribution $ P_s(C) $, but only the transition rates $W[C \to C']$ are known. Thus, we need to find out $ P_s(C) $ from the transition rates without assuming that the DB condition holds a priori; this is usually a very challenging task. However, it is easy to prove~\cite{Kolm36,Zia2007} the following equivalent condition known as Kolmogorov's criterion: if the DB condition is satisfied, then for every set of configurations $\{ C_1, C_2, \dots , C_n \}$:
\begin{equation}
\displaystyle
\begin{split}
& W[C_1 \to C_2] W[C_2 \to C_3] \dots  W[C_n \to C_1]  = \\
& W[C_n \to C_{n-1}] \dots W[C_2 \to C_1] W[C_1 \to C_n] \ .
\end{split}
\label{eq:DBcicle}
\end{equation}
This equation is a condition that depends only on the transition rates and it always holds for equilibrium steady states. It is also a time reversal symmetry property: given a configuration of the system, the probability to come back to the same configuration following a path in the space of configurations does not depend on the orientation of the path itself. Thus, eq.~\ref{eq:DBcicle} suggests a simple scheme for assessing whether the OG model has non-equilibrium characteristics: we only need to find a set of configurations which do not satisfy the condition in eq.~\ref{eq:DBcicle}.

\begin{figure}[tbp]
\centering
  \includegraphics[width=0.6\textwidth]{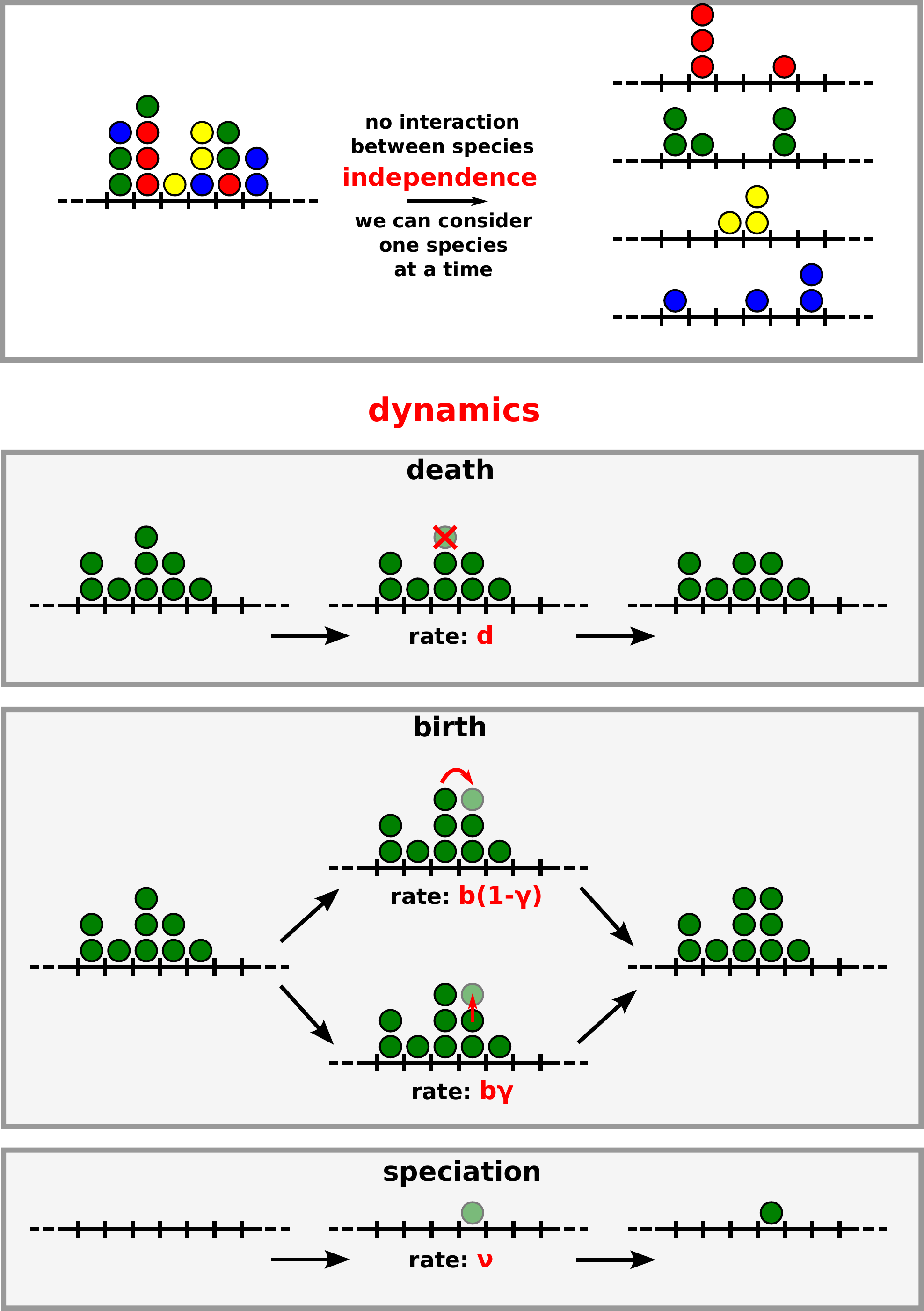}
	\caption{Definition of the OG model. The model is non-interacting (different species are totally decoupled) and neutral (the parameters of
	the model do not depend on the species we are considering). Because of these assumptions
	it is possible to consider one species at a time. In the figure, a one dimensional
	lattice is shown for simplicity. A circle corresponds to a single individual
	and its color represents the species it belongs to. There is no limit to the number of individuals that can be placed in each site.
	The dynamics is defined in terms of three possible main moves: birth, death and speciation.
	The per capita death rate is equal to $d$, whereas the speciation rate (defined per unit of lattice
	spacing) is $\nu$. The total birth rate is equal to $b$: with a probability $\gamma$, the offspring
	is placed in the same site as the parent, while, with a probability $1 - \gamma$, it is
	placed in a different site chosen uniformly among the neighboring sites (which corresponds to a dispersal move).}
\label{fig:modeldef}
\end{figure}

\begin{figure}[tbp]
\centering
  \includegraphics[width=0.6\textwidth]{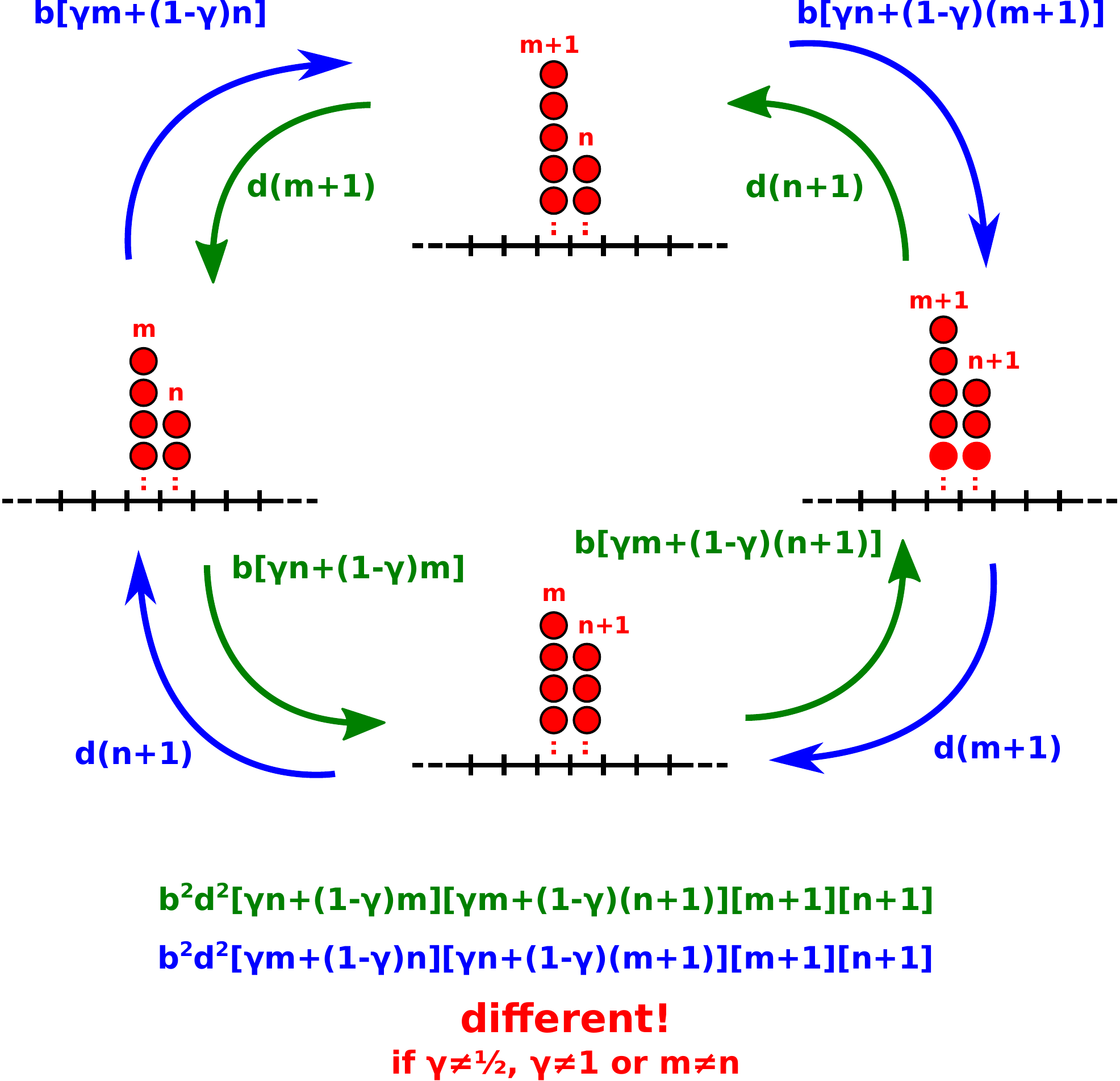}
  \caption{Detailed Balance in the OG model. In general, detailed balance is not satisfied if the probability to go along a closed path in the space of configurations depends on the direction that one chooses (see eq.~\ref{eq:DBcicle}). Here we present the simplest counterexample to eq.~\ref{eq:DBcicle} for the OG model. The probability to find the system in any of the four configurations in red and linked by the cycle depends on whether one follows the green or blue path. When $\gamma=1$, detailed balance is valid, whereas, for $\gamma=1/2$, it is generally not valid (i.e. even though the condition of eq.~\ref{eq:DBcicle} is valid for this particular cycle, it is not valid generally).}
\label{fig:modeldb}
\end{figure}

The OG model~\cite{O'Dwyer2010} is a birth-death-dispersal neutral process~\cite{Hubbell2001a} defined on a regular lattice $\Lambda$ (see fig.~\ref{fig:modeldef}). The configurations of the model are represented by the number of individuals, which are present on each site. The dynamics of different species are totally decoupled, i.e., there are no interactions between individuals of different species~\cite{Volkov2003,Volkov2009,Sandro10}. At any given time $t$, a species has a total number of individuals $n(t)$ and any individual can die at a rate $d$ or survive and produce one offspring at a rate $b$ if there is at least one individual in the lattice, i.e., $n > 0$. When the species population is zero everywhere, i.e., $n = 0$, a speciation/immigration event can occur with speciation rate per site equal to $\nu$. The non-trivial spatial dependence of the model is given by the birth event: when an individual in the site $i$ produces an offspring, the new individual remains in the same site with probability $\gamma$ or, alternatively, is located within one of the other sites with probability $1-\gamma$. The model can be formulated with a general dispersion kernel, that defines the probability that the new offspring is located in a site $j$ given a parent in a site $i$. The model was solved only in the case of dispersion to nearest neighbor sites and we focus our analysis on this case. All the consequences we present are also valid in the more general case.
Thus, the dynamics of the model is completely specified by the following transition rates which define the master equation via eq.~\ref{eq:mastereq}
\begin{equation}
\displaystyle
\begin{split}
 W[ \{n_1, \dots , n_i, \dots, n_{|\Lambda|} \} \to \{n_1, \dots , n_i + 1, \dots, n_{|\Lambda|} \}] & =
b \gamma n_i + b \frac{1-\gamma}{\mu} \sum_{j:|j-i|=1} n_j + \nu \prod_{j=1}^{|\Lambda|} \delta_{n_j,0}  \\
 W[ \{n_1, \dots , n_i, \dots, n_{|\Lambda|} \} \to \{n_1, \dots , n_i - 1, \dots, n_{|\Lambda|} \}] & =
d n_i  ,
\end{split}
\label{eq:Transrate}
\end{equation}
where $|\Lambda|$ is the total number of sites in the lattice and $n_i$ is the population on the site $i$.
As we said, it is sufficient to find a set of configuration
which do not satisfy the equation~\ref{eq:DBcicle} to show that the DB does not hold. Figure~\ref{fig:modeldb}
shows a set of configuration which, given the rates defined in eq.~\ref{eq:Transrate},
does not respect the reversibility condition of eq.~\ref{eq:DBcicle} and therefore the DB condition is not valid for the model. When $\gamma=1$ there
is not dispersion in space, and the model becomes a simple birth-death process for which DB holds.

If a model  satisfies DB, the spatio-temporal evolution following a speciation event with time running forwards can be interpreted (on average) as the spatio-temporal evolution preceding an extinction event with time running backwards. Thus the processes after speciation and before extinction events are temporally symmetric. For the OG model, instead, speciation and extinction are temporally asymmetric processes (as is observed in nature): the spatio-temporal behavior of species that follows a speciation event is completely different (on average) from the spatio-temporal behavior that precedes an extinction event.

By introducing the moment generator function $Z$, defined as
\begin{equation}
\displaystyle
Z(\underline{h},t) = \sum_{n_1,\dots,n_{|\Lambda|}} P(n_1, \dots, n_{|\Lambda|}; t) e^{\sum_{i \in \Lambda} n_i h_i}  \ ,
\label{eq:partfunc}
\end{equation}
we obtain the following evolution equation:
\begin{equation}
\displaystyle
 \frac{\partial Z(\dots, h_i, \dots, t)}{\partial t} =
 \sum_{i \in \Lambda} \big( e^{h_i}-1 \big) \Bigl[ \big( b \gamma - d e^{- h_i} \big)
\frac{ \partial Z }{ \partial h_i }
+ b \frac{1-\gamma}{\mu} \sum_{j:|j-i|=1} \frac{ \partial Z }{ \partial h_j }
+ b \tilde{\theta}(t) \Bigr]  \ ,
\label{eq:DwyerZ1}
\end{equation}
where $\tilde{\theta}(t)$ is defined as $\nu P(n_1=0,n_2=0,\dots,n_{|\Lambda|}=0;t) / b $ and
$\mu$ is the number of neighbors in the lattice (e. g. in a $D$ dimensional square lattice $\mu=2 D$).
We can introduce in the previous expression the discrete Laplacian operator $\Delta_\Lambda$
defined over a generic regular lattice $\Lambda$ with periodic boundary conditions
\begin{equation}
\displaystyle
(\Delta_\Lambda f )_i = \sum_{j:|j-i|=1} \big( f_j - f_i \big) \ ,
\label{eq:LapDef}
\end{equation}
obtaining
\begin{equation}
\displaystyle
 \frac{\partial Z(\dots, h_i, \dots, t)}{\partial t} =
 \sum_{i \in \Lambda} \big( e^{h_i}-1 \big) \Bigl[ \big( b - d e^{- h_i} \big)
\frac{ \partial Z }{ \partial h_i }
+ b \sigma^2 \Bigl( \Delta_\Lambda \frac{ \partial Z }{ \partial h } \Bigl)_i
+ b \tilde{\theta}(t) \Bigr]  \ ,
\label{eq:DwyerZ2}
\end{equation}
where $\sigma^2=(1-\gamma)/\mu$.
In the continuum limit (see the Supplementary Materials of~\cite{O'Dwyer2010}),
the discrete Laplacian $\Delta_\Lambda$ becomes the Laplacian operator $\Delta$ and
thus we obtain
\begin{equation}
\displaystyle
\frac{ \partial \mathscr{Z}[H,t] }{\partial t} = \int_{A_0} d^D\underline{x} \big( e^{H(\underline{x})} -1 \big)
\Bigl[ b \sigma^2 \Delta \frac{ \delta \mathscr{Z}[H,t] }{ \delta H(\underline{x}) }
+ \big( b - d e^{-H(\underline{x})} \big) \frac{ \delta \mathscr{Z}[H,t] }{ \delta H(\underline{x}) }
+  b \tilde{\theta}_s(t) \Bigr] \ ,
\label{eq:DwyerMastGen}
\end{equation}
where $A_0$ is the area (volume) of the entire system.

One can show that the solution proposed by the authors in ref.~\cite{O'Dwyer2010} is equivalent
to imposing DB. This condition for the OG model (see fig.~\ref{fig:modeldef}) can be written as
\begin{equation}
\begin{split}
\displaystyle
& d (n_i+1) P_s^{DB}(\dots, n_i+1, \dots,n_j,\dots) = \\
& \Bigl[ b \gamma n_i + b \frac{1-\gamma}{\mu} \sum_{j:|j-i|=1} n_j + (\prod_j \delta_{n_j,0}) \nu \Bigr]
 P_s^{DB}(\dots, n_i, \dots,n_j,\dots)
\label{eq:DwyerDetBal}
\end{split}
\end{equation}
for any choice of $i$.
The notation $P_s^{DB}$ stands for the stationary probability (i.e. the solution of the right hand side
of master equation set equal to zero) obtained by imposing the DB condition.
By introducing into this expression the definition of $Z$, we obtain:
\begin{equation}
\begin{split}
\displaystyle
d e^{h_i} \frac{ \partial Z_s^{db} }{\partial h_i} =
 b \gamma \frac{ \partial Z_s^{db} }{\partial h_i} + b \sigma^2 \sum_{j:|j-i|=1}
\frac{ \partial Z_s^{db} }{\partial h_j} + b \tilde{\theta}  \ ,
\label{eq:DwyerDetBal}
\end{split}
\end{equation}
where $\tilde{\theta}$ is now equal to $\nu P_s^{DB}(n_1=0, \dots, n_{|\Lambda|}=0)/b$.
This expression, in the continuum limit, becomes
\begin{equation}
\displaystyle
b \sigma^2 \Delta \frac{ \delta \mathscr{Z}^{db}[H,t] }{ \delta H(\underline{x}) }
+ \big( b - d e^{-H(\underline{x})} \big) \frac{ \delta \mathscr{Z}^{db}[H,t] }{ \delta H(\underline{x}) }
+ b \tilde{\theta}_s = 0 \ .
\label{eq:DwyerDetBalZ}
\end{equation}
This is exactly the equation solved in ref.~\cite{O'Dwyer2010} .
Thus we have shown that the solution obtained by the authors by considering eq.~\ref{eq:DwyerDetBalZ} instead of
the right side of eq~\ref{eq:DwyerMastGen}
corresponds to imposing the detailed balance condition which is not valid for the OG model
(as shown in fig.~\ref{fig:modeldb}).

Assuming DB to hold when it is violated produces incorrect results. An example arises in the $l$-point correlation function. OG obtained an equation for the moment generating function $Z$ defined in eq.~\ref{eq:partfunc}: by expanding it, one can obtain the equation for the $l$-points correlation function, $\big< n_{i_1} n_{i_2} \dots n_{i_l} \big>:=G_l(i_1,\dots,i_l)$. Because of its definition, the correlation function must be symmetric under the exchange of its arguments,
e.g., for the $3$-point correlation function $G_3(i_1,i_2,i_3)=G_3(i_2,i_1,i_3)$.

We can show that applying DB to the OG model (i.e. by solving
equation~\ref{eq:DwyerDetBalZ}), one obtains asymmetric correlation functions. Indeed the $3$-point correlation function
$G_3(i_1,i_2,i_3)$ does \emph{not} possess the required symmetry.
To see this in an explicit way, a useful procedure to simplify and solve the involved differential equations is to apply the Fourier transform and consider the function $\widehat{G}_3(\underline{p}_1,\underline{p}_2,\underline{p}_3)$.
Because of the properties of the Fourier transform, we know that the function $G_3$ is symmetric in its
arguments if and only if $\widehat{G}_3$ is symmetric. By considering the first three correlation functions,
we obtain the following expression:
\begin{equation}
\begin{split}
\displaystyle
\widehat{G}_1(\underline{p}_1) = & \frac{b \tilde{\theta}}{b-d} \delta(\underline{p}_1) = \big<n\big> \delta(\underline{p}_1) \\
\widehat{G}_2(\underline{p}_1,\underline{p}_2) = &
\frac{ - d \widehat{G}_1(\underline{p}_1+\underline{p}_2)}{- b \sigma^2 \underline{p}_1^2 + (b-d)}
= \frac{- d \big<n\big>}{- b \sigma^2 \underline{p}_1^2 + (b-d)}\delta(\underline{p}_1+\underline{p}_2) \\
\widehat{G}_3(\underline{p}_1,\underline{p}_2,\underline{p}_3) = &
d \frac{ \widehat{G}_1(\underline{p}_1+\underline{p}_2+\underline{p}_3)-\widehat{G}_2(\underline{p}_1+\underline{p}_2,\underline{p}_3)-
\widehat{G}_2(\underline{p}_1+\underline{p}_3,\underline{p}_2)}{- b \sigma^2 \underline{p}_1^2 + (b-d)} = \\
= & \frac{2 d \big<n\big> - b \sigma^2 \underline{p}_1^2 + (b-d) }{( b \sigma^2 \underline{p}_1^2 - b+d)^2}
\delta(\underline{p}_1+\underline{p}_2+\underline{p}_3) \ .
\end{split}
\label{eq:Dwyer3CorWrong}
\end{equation}
The three point correlation function is not symmetric in its arguments.
It clearly shows that, on assuming DB in the OG model, we obtain
inconsistencies in the solution. This inconsistence is valid for all the $l$-point correlation functions, except the $2$-point correlation function.

\begin{figure}[tbp]
\centering
  \includegraphics[width=0.9\textwidth]{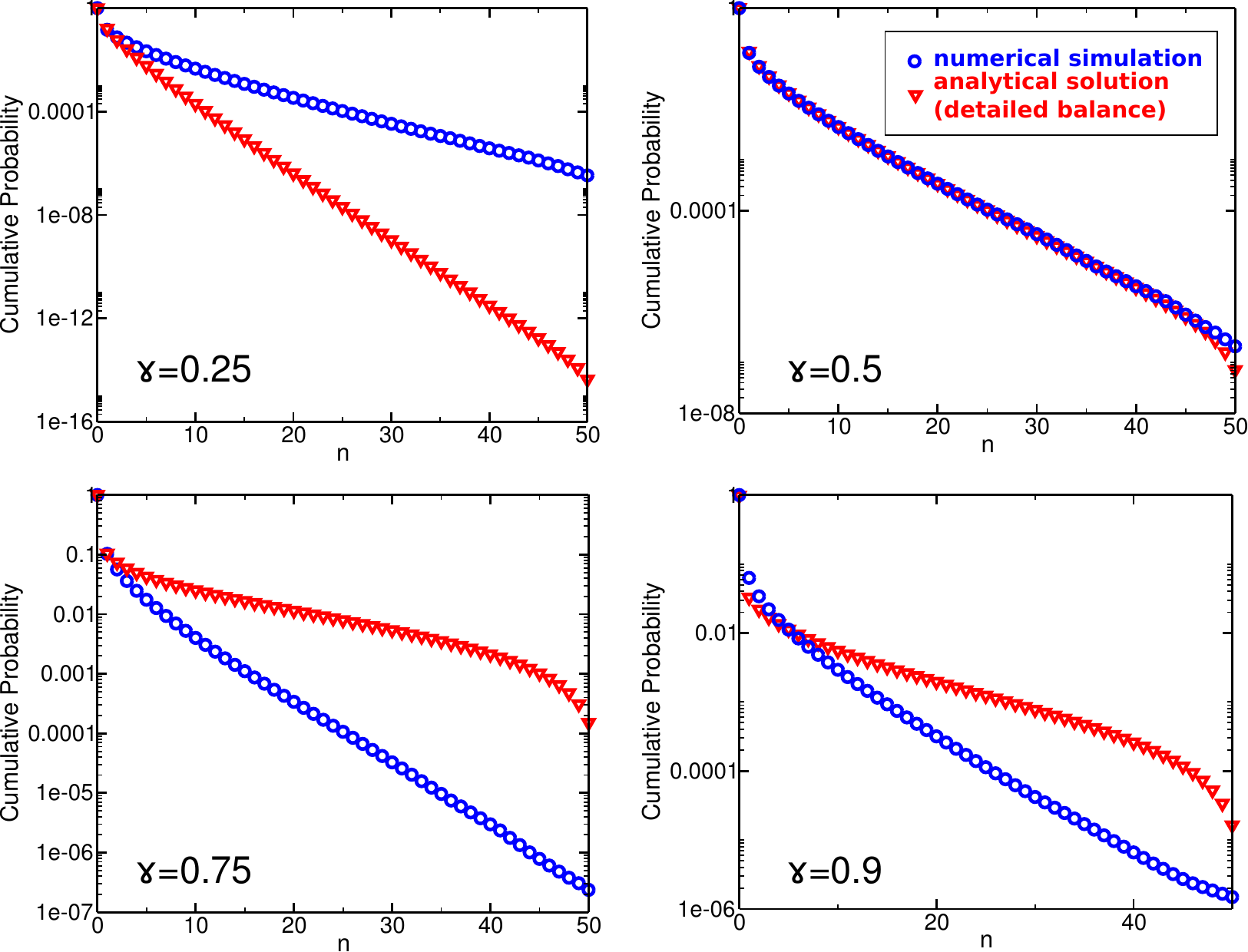}
  \caption{Comparison between the simulations of the OG model and the analytical solution obtained
	by applying the DB condition. The blue circles are simulations of
	the cumulative stationary probabilities $P_1^{>}(n)=\sum_{k=n}^\infty\sum_{m=0}^\infty P(k,m)$.
	The red triangles represents $P^{>DB}_1(n)$, obtained  by summing over $m$
	eq.~\ref{eq:2sitesDB-res}, which is calculated by imposing DB.
	As expected, the solutions are different, except in the
	case $\gamma=1/2$ (as expected from fig.~\ref{fig:modeldb}).
	For this value of $\gamma$ the DB is still not valid ($3$-point correlation function is still non-symmetric).	
	This special behavior of the model for $\gamma=0.5$ is therefore caused
	by the smallness of the system considered in this case,
	and not because DB is restored.
	The model is simulated with the following parameters: $b=0.9$, $d=1.$ and $\nu=0.1$.}
\label{fig:simuldb}
\end{figure}

We have also simulated the model in a simple case: we show in fig.~\ref{fig:simuldb} that the analytical results do not match the numerical simulations of the model. We consider the OG with just two sites.
It is the simplest case in which DB is not valid (i.e.
it is possible to construct a cycle of configurations as in fig.~\ref{fig:modeldb}).
We find a stationary probability by applying DB and show that this condition
produces inconsistencies.

Consider the generic configuration $(n,m)$. The DB conditions can be expressed as
\begin{equation}
\begin{cases}
\displaystyle
d P^{DB}(1,0) = \frac{\nu}{2}  P^{DB}(0,0)  \\
d (n+1) P^{DB} ( n+1,m) = b \big[ \gamma n + (1-\gamma) m \big] P^{DB}(n,m)  \ .
\end{cases}
\label{eq:2sitesDB}
\end{equation}
We can thus simply construct a solution by imposing DB. Specifically, we obtain for $m$ and $n$ bigger than zero
\begin{equation}
\displaystyle
P^{DB}(n,m) = \tilde{\theta} \frac{(x \gamma)^{n+m}}{m n} \Biggl[
\frac{ \Gamma \big( n + \frac{1-\gamma}{\gamma}m \big)}{ \Gamma \big( n \big) \Gamma \big( \frac{1-\gamma}{\gamma} m \big) }
\Biggr] \ ,
\label{eq:2sitesDB-res}
\end{equation}
where $\Gamma(\cdot)$ is the Gamma function and $x=b/d$.
This expression is clearly inconsistent because it is not symmetric under the exchange of its arguments $(n,m)$.
We can see that this expression turns to be consistent for $\gamma = 1$ and $\gamma=1/2$, exactly the choices of
parameters for which the reversal symmetry in fig.~\ref{fig:modeldb} is restored. This inconsistency is shown in
fig.~\ref{fig:simuldb}, where we compare the analytical result with the numerical simulations.
The case $\gamma=1$ (which corresponds to $\sigma=0$) is trivial because in this case we are not considering
diffusion in space and the sites are independent (see fig.~\ref{fig:modeldef})~\cite{Volkov2003},
and so the DB condition is valid. In fact the $3$-point correlation
function in eq.~\ref{eq:Dwyer3CorWrong} is symmetric in the case $\gamma=1$. When we consider $\gamma=1/2$ we expect the
DB not to be valid, because the $3$-point correlation function is still non-symmetric. The fact that the DB seems to be valid
in the simulation of fig.~\ref{fig:simuldb} is simply due to the fact that we are considering a very simple case with only two sites.

We have shown that the solution obtained by applying the DB is not the real stationary solution of the model and that one obtains
incorrect results. It is not possible to treat that solution as an approximation to the correct one, because there is no parameter which is able to quantify the goodness of the approximation. In addition, the 3-point correlation function obtained on assuming DB is not symmetric and this leads to incorrect results, regardless of the nature of the approximation.

One may ask whether the formula obtained by applying the DB could provide a good
fit to the data. Unfortunately this is not the case. It is known that, at very small areas, the SAR grows
linearly with the area~\cite{Hubbell2001a,jtb_sar} and, in particular, it is equal to the number of sampled individuals. By expanding at small areas ($r \ll \sigma$) the solution obtained by imposing the DB, we obtain:
\begin{equation}
\displaystyle
S(r) \sim N(r) \log \Bigl( \frac{\sigma^2}{\alpha r^2} \frac{2}{(1+\alpha)
( \log(\frac{2 \sigma}{\alpha r}) -\gamma_E )  }  \Bigr)   \ ,
\label{eq:sar-exp}
\end{equation}
where $\gamma_E$ is the Euler-Mascheroni constant ($\gamma_E \approx 0.577 \dots$) and $N(r)=\frac{\tilde{\theta}}{\alpha} \pi r^2$ is the number of individuals sampled. Note that this expression is not close to the expected linear growth $S(r) \sim \rho \pi r^2 = N(r)$ (where $\rho$ is the density of individuals), because the logarithm is much larger than one ($\sigma/r \gg 1$). Moreover, the expansion leads to the meaningless result that the number of species sampled is larger than the number of individuals.
This wrong prediction is one of the consequences of the use of the DB condition.

The OG model was recently proposed~\cite{O'Dwyer2010} to describe the stationary properties of an ecosystem. This model appeared to be the first solvable model able to predict the empirical Species-Area Relationship, which is one of the most important stationary quantities in ecology. OG solved the model in an elegant manner via the moment generator functional. We have shown that the OG model is not an equilibrium model, as demonstrated in fig.~\ref{fig:modeldb}.
The OG model is a birth-death-migration process: a species changes its population and the individuals diffuse in space, a process which is not reversible. Furthermore, there are inconsistencies in the solution on wrongly assuming detailed balance such as the correlation functions no longer being symmetric in their arguments. The basic message is that the behaviors and the properties of the stationary state of equilibrium and non-equilibrium systems can be distinct.
Life is not an equilibrium system nor it is reversible. Developing techniques for studying the behavior of models, which do not obey detailed balance, is a necessary but daunting step to understand the physics of non-equilibrium processes. An alternative to the OG model based on the Poisson cluster processes, which is solvable and leads to an analytic expression for the SAR, has been recently developed~\cite{jtb_sar}.

\acknowledgments
JG and AM thank Cariparo foundation for financial support.

\bibliographystyle{unsrt}
\bibliography{biblio}

\end{document}